**A Soft Sponge-Pillar Electrode Platform for Motion-Tolerant Ambulatory EEG and Sleep Monitoring**

*Chansoo Kim†, Weilun Li†, Ryan Andersen, Yiheng Zhang, Shuoyan Liu, Junyi Zhao, Yuelin Li, Zhongyu Luo, Josephine Lau, Yang Huang, Diane Bracken, Matthew Singh, ShiNung Ching, Ben Julian A. Palanca, Yong Wang, Shantanu Chakrabartty, and Chuan Wang**

Chansoo Kim, Weilun Li, Ryan Andersen, Yiheng Zhang, Shuoyan Liu, Junyi Zhao, Yuelin Li, Matthew Singh, ShiNung Ching, Shantanu Chakrabartty, and Chuan Wang
Electrical & System Engineering, Washington University in St. Louis, St. Louis, Missouri, United States of America
E-mail: chuanwang@wustl.edu

Chansoo Kim and Chuan Wang
Institute of Materials Science and Engineering, Washington University in St. Louis, St. Louis, Missouri, United States of America

Zhongyu Luo
Image Science, Washington University in St. Louis, St. Louis, Missouri, United States of America

Josephine Lau
Biomedical Engineering, Washington University in St. Louis, St. Louis, Missouri, United States of America

Diane Bracken and Ben Julian A. Palanca
Anesthesiology, Washington University School of Medicine, St. Louis, Missouri, United States of America

Zhongyu Luo, Josephine Lau, Yuelin Li, Yang Huang, and Yong Wang
Obstetrics and Gynecology, Washington University School of Medicine, St. Louis, Missouri, United States of America

† Chansoo Kim and Weilun Li contributed equally to this work.

**Abstract**
Long-term wearable electrophysiological recording systems are essential for both clinical and real-world environments. While a wide range of wearable devices have been reported for electrocardiogram (ECG) and electromyogram (EMG) recording applications, most electroencephalography (EEG) measurements still require subjects to remain stationary while wearing rigid and uncomfortable equipment. Here, a soft and wearable sponge pillar EEG electrode is presented for long-term and motion-tolerant EEG recording. The electrode is composed of two layers of a carbon black (CB)-polydimethylsiloxane (PDMS) composite, where a firmer solid base assists in bypassing hair, enabling stable access to the scalp, while a porous sponge tip ensures comfort, reliable contact, and low skin-electrode impedance. Fabricated via a sugar templating process without microfabrication steps, the microporous sponge structure offers low modulus and significantly enhances the contact area with the skin when loaded with conductive gel. This allows the electrode to achieve a lower contact impedance of $1.62\times10^4$ Ω at 10 Hz, outperforming conventional gelled electrodes while maintaining gel stability for over 18 hours. With the electrodes integrated into wearable form factors such as headbands, headphones, and baseball caps, EEG signal recording during motion have been demonstrated, achieving a signal-to-noise ratio (SNR) of 16.46 dB. These features establish the sponge pillar electrode as a promising platform for next generation wearable EEG systems, enabling comfortable, stable, and prolonged brain signal acquisition on ambulatory subjects and in naturalistic environments.

## 1. Introduction

With the growing demand for continuous and real-time health monitoring, wearable systems have rapidly emerged as essential tools for next generation healthcare, with broad applications ranging from clinical diagnostics to personal wellness management. Especially, systems capable of non-invasively capturing bioelectrical signals from the skin surface enable comfortable, long-term monitoring across diverse environments.[1, 2] Among various bio signals, electrocardiography (ECG) plays a critical role for predicting arrhythmia,[3, 4] myocardial infarction,[5, 6] and left ventricular hypertrophy[7] and electromyography (EMG) is widely utilized in diagnosing muscular dystrophy,[8, 9] myasthenia gravis,[10, 11] and peripheral nerve disorders,[12, 13] by acquiring electrical signals from the epidermal surface. Electroencephalography (EEG), which records brain activity with high temporal resolution,[14] has proven indispensable in neuroergonomics,[15, 16] clinical diagnosis,[17-19] sleep medicine,[20, 21] anesthesia monitoring,[22, 23] and brain computer interfaces (BCI).[24, 25]

Despite its wide applicability, EEG remains one of the most challenging signals to acquire in wearable settings. In contrast to ECG and EMG, EEG signals are inherently low in amplitude (typically <100 μV), making them especially vulnerable to noise and motion artifacts.[26] In addition, the presence of hair significantly obstructs stable skin contact, further increasing impedance and degrading signal quality.[16] Commercial EEG electrodes attempt to overcome this challenge by employing comb like or pillar structures designed to bypass hair or small cup shaped electrodes to retain conductive gel at the contact interface. However, these designs are often rigid and non-conformal, compromising user comfort and stability during motion. In addition, gel-based electrodes suffer from signal degradation over time due to gel drying.[27] These limitations hinder long-term usability and reduce practicality in dynamic conditions. To address these challenges, various strategies have been proposed to improve the stability and fidelity of wearable EEG acquisition. These include the development of mechanically compliant electrode structures that better conform to the scalp,[28, 29] microneedle arrays that penetrate the stratum corneum to reduce impedance between skin and electrode,[30] and adhesive layers that enhance contact stability.[31, 32] While each approach demonstrates potential benefits, many of these systems require complex microfabrication processes, limiting their scalability and increasing production cost. Furthermore, most existing designs assume a stationary user during recording, restricting their applicability to real-world, ambulatory monitoring.

In this work, we introduce a soft, flexible EEG electrode composed of a carbon black (CB) and polydimethylsiloxane (PDMS) composite in a sponge pillar structure (**Figure 1a**). This design facilitates effective penetration through hair to reach the scalp surface while reliably delivering ionic conducting gel and maximizing the contact area.[27] At the same time, motion artifacts are minimized through mechanical compliance. Owing to these structural advantages, the electrode achieved lower contact impedance compared to conventional commercial EEG electrodes. Building on its structural advantages, the electrode was implemented across multiple wearable form factors including headbands, headphones, and baseball caps and it demonstrated reliable signal acquisition even during movement. Its design was further extended by integration with a screen-printed textile patch using poly(3,4-ethylenedioxythiophene):poly(styrene sulfonate) (PEDOT:PSS), enabling multichannel EEG mapping and supporting feasibility for long-term and dynamic brain monitoring applications. Additionally, its comfort and softness enabled successful application in long-duration sleep monitoring, consistently capturing high-quality EEG signals. Overall, these results highlight the potential of organic and soft materials to simplify and broaden the design of wearable bioelectronic systems, offering a scalable and clinically relevant platform for continuous brain health monitoring in naturalistic settings.

## 2. Results

### 2.1. Sponge pillar structure

The core structural design of the developed EEG electrode is illustrated in Figure 1a. The electrode adopts a vertically stacked two-layer architecture consisting of a soft, porous sponge tip and a rigid, non-porous conductive base. This design is intended to address common challenges in EEG acquisition, such as achieving reliable scalp contact through dense hair while minimizing motion induced artifacts. In addition, to enhance the system's modularity and adaptability for wearable applications, the sponge pillar electrode is designed to be integrated with a textile-based electrode platform fabricated via screen printing of PEDOT:PSS (**Figure 1b**). A key aspect of the electrode design lies in the rationale behind the two-layer pillar structure (**Figure 1c**). The rigid upper layer provides structural support to navigate hair and ensure positional stability. Meanwhile, the lower sponge portion features a compliant and porous structure that increases the contact surface area with both the ionic gel and the skin, thereby significantly reducing the skin-electrode impedance. The softness of the sponge also helps to buffer external forces and mitigate motion artifacts during dynamic activities. The mechanical benefits of this design are clearly demonstrated in **Figure 1d**, which compares the flexibility between the sponge pillar electrode and a control electrode composed of a solid PDMS-based pillar. The solid pillar shows noticeable rigidity, failing to deform in response to applied forces, which could lead to pressure points and unstable contact under movement. In contrast, the sponge pillar electrode exhibits excellent compliance, naturally bending along the direction of force, maintaining stable skin contact, and improving wearing comfort. The applicability of this electrode system is further illustrated in **Figure 1e**, where the electrodes are implemented into various wearable form factors, including headbands, headphones, and baseball caps. These configurations demonstrate the system's robust performance in both stationary and dynamic conditions, including daily activity monitoring and overnight sleep EEG recording. The ability to maintain high-quality signal acquisition across different wearables highlights the practicality and scalability of the proposed system for long-term, real-world electrophysiological monitoring.

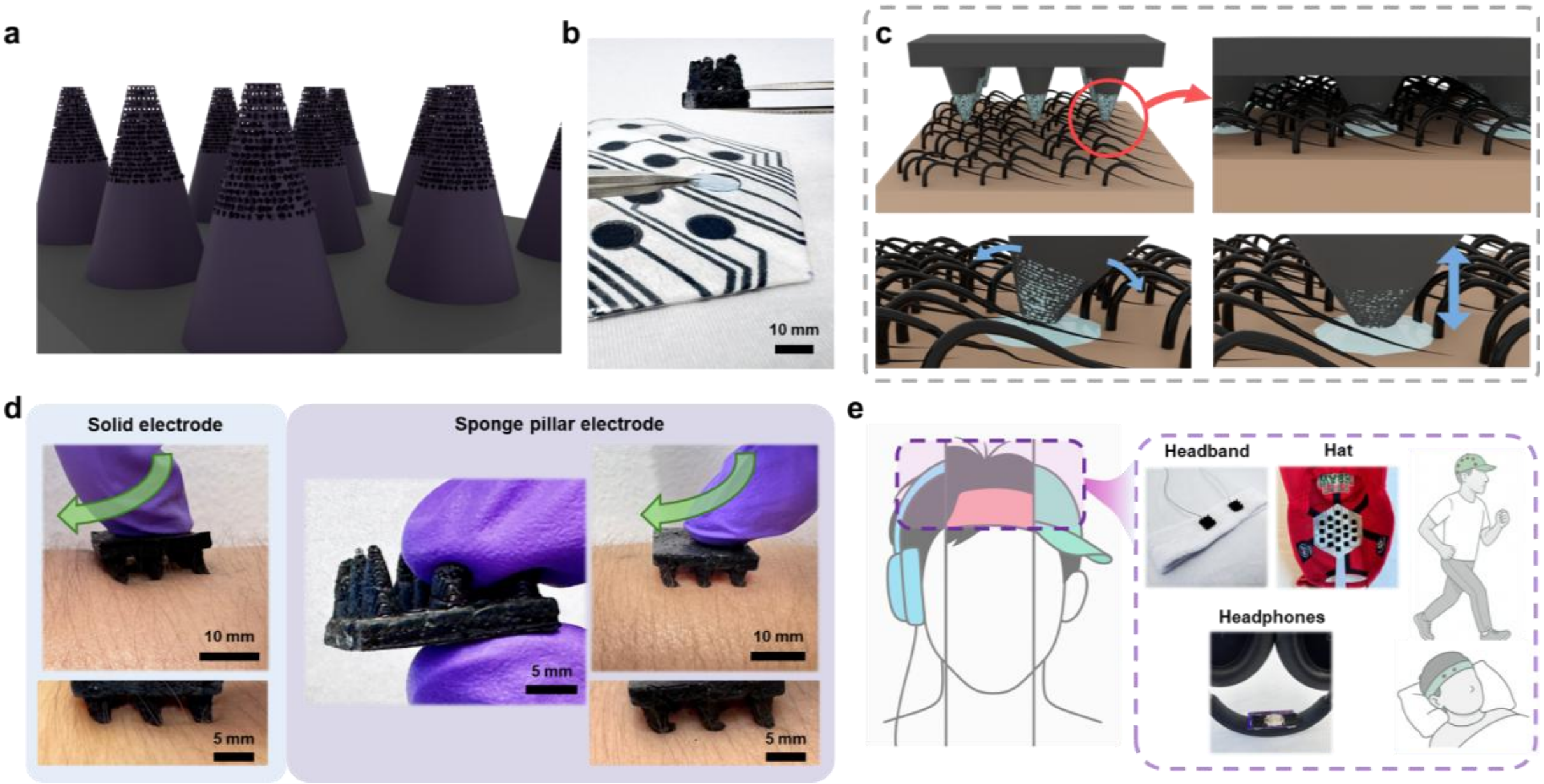

**Figure 1.** Illustration of sponge pillar electrode. a) Schematic illustration of the sponge pillar electrode composed of a two-layer structure, including a porous CB-PDMS sponge tip and a rigid base layer. The sponge tip enhances flexibility and enlarges the effective contact area, while the rigid base provides sufficient stiffness to penetrate through hair. b) The picture showing integration of the sponge electrode with a PEDOT:PSS screen-printed textile electrode. An ionic gel binder is used to provide a conformal and stable interface between the textile and sponge electrode. c) Conceptual diagram of the electrode-scalp interface. The porous sponge tip delivers ionic gel to the scalp while minimizing exposure to air, which supports gel retention and reduces evaporation. The increased surface area from the internal

pores also enhances skin-electrode contact. Additionally, the mechanical flexibility of the sponge layer helps to maintain stable contact during movement. d) Comparison of mechanical behavior between a solid CB-PDMS electrode and the sponge pillar electrode during skin contact. The rigid structure fails to accommodate movement, resulting in potential discomfort and signal instability. In contrast, the sponge electrode exhibits superior compliance under compression and shear, allowing it to maintain stable contact under motion. e) Application of the electrode across multiple wearable form factors, including headbands, headphones, and baseball caps, enabling EEG acquisition during ambulation and sleep.

### 2.2. Electrode properties

**Figure 2a** and **Figure S1** present the macroscopic and microscopic images of the porous sponge pillar electrode. The images clearly demonstrate the distinct two-layer structure, where the base consists of a solid, non-porous PDMS-carbon black composite, and the tip features a porous sponge microstructure. The detailed fabrication process is schematically illustrated in **Figure 2b**. In contrast to conventional EEG electrodes that rely on expensive conductive materials such as gold, the proposed fabrication method eliminates the need for costly metals, resulting in electrodes that are softer, more compliant, and more cost effective than commercial counterparts. These characteristics significantly improve user comfort, particularly for long-term wear during daily activities. Additionally, the simplicity of the fabrication process allows for scalable production and supports potential use as disposable electrodes.

The fabrication begins with the preparation of two distinct precursor mixtures: (1) a mixture of PDMS, carbon black (CB), and sugar for the sponge tip, and (2) a mixture of PDMS and CB for the rigid base. To form the porous sponge structure, half of the cavity in a 3D printed mold is filled with the sugar-containing mixture, which is then compacted manually using a rod to ensure uniform packing and structural integrity, as described in the Methods section. Subsequently, the upper half is sequentially filled with the PDMS-CB mixture to form the solid base. The assembled mold undergoes a thermal curing process to solidify the PDMS matrix, followed by a sugar dissolution step to create the porous architecture in the tip, resulting in the final sponge pillar EEG electrode.

The electrical performance of the sponge pillar electrode was evaluated by measuring the skin and electrode contact impedance and comparing it with commercial gold and comb EEG electrodes (**Figure 2c**). Under gel-based conditions, the sponge pillar electrode demonstrated the lowest impedance of $1.62\times10^4$ Ω at 10 Hz, compared to $4.57\times10^4$ Ω for the gold electrode and $3.24\times10^4$ Ω for the comb. This superior performance is attributed to the increased effective contact area provided by the interconnected micropores within the sponge structure, which enhances ionic gel coupling with the skin. Furthermore, the porous architecture not only lowers impedance but also plays a critical role in maintaining gel stability over time. The internal pore structure minimizes the contact between the gel and air, effectively reducing evaporation and prolonging gel retention. This property was evaluated through a long-term impedance test over 18 hours under ambient air conditions (**Figure 2d**).[27] During the test, all electrodes were exposed to air after the initial application of the ionic gel. The contact impedance gradually increased from $2.77\times10^4$ Ω to $5.82\times10^4$ Ω, with the rate of increase plateauing after approximately 4 hours, indicating stabilized gel retention beyond that point. Consistent with the impedance results, long-term EEG recordings further confirmed stable signal acquisition. As shown in **Figure S2**, the alpha wave signal elicited by eye closure was continuously monitored at multiple time points over the 18-hour period. The calculated signal-to-noise ratio (SNR) remained stable within the range of 12 to 15 dB, indicating no significant signal degradation over time despite prolonged exposure to air. The mechanical advantage of the sponge structure is further demonstrated in **Figure 2e**, which presents the tensile test results comparing the sponge pillar electrode to a solid PDMS-CB electrode. The solid electrode exhibited a Young's modulus of 164.98 MPa, whereas the sponge electrode achieved a significantly lower modulus of 43.55 MPa. The corresponding standard deviations were 0.62 MPa for the solid electrode and 10.38 MPa for the sponge electrode. This enhanced softness allows the sponge pillar electrode to better accommodate skin movement and external forces, thereby reducing motion induced artifacts and improving overall comfort during EEG recording (Figure 1d).

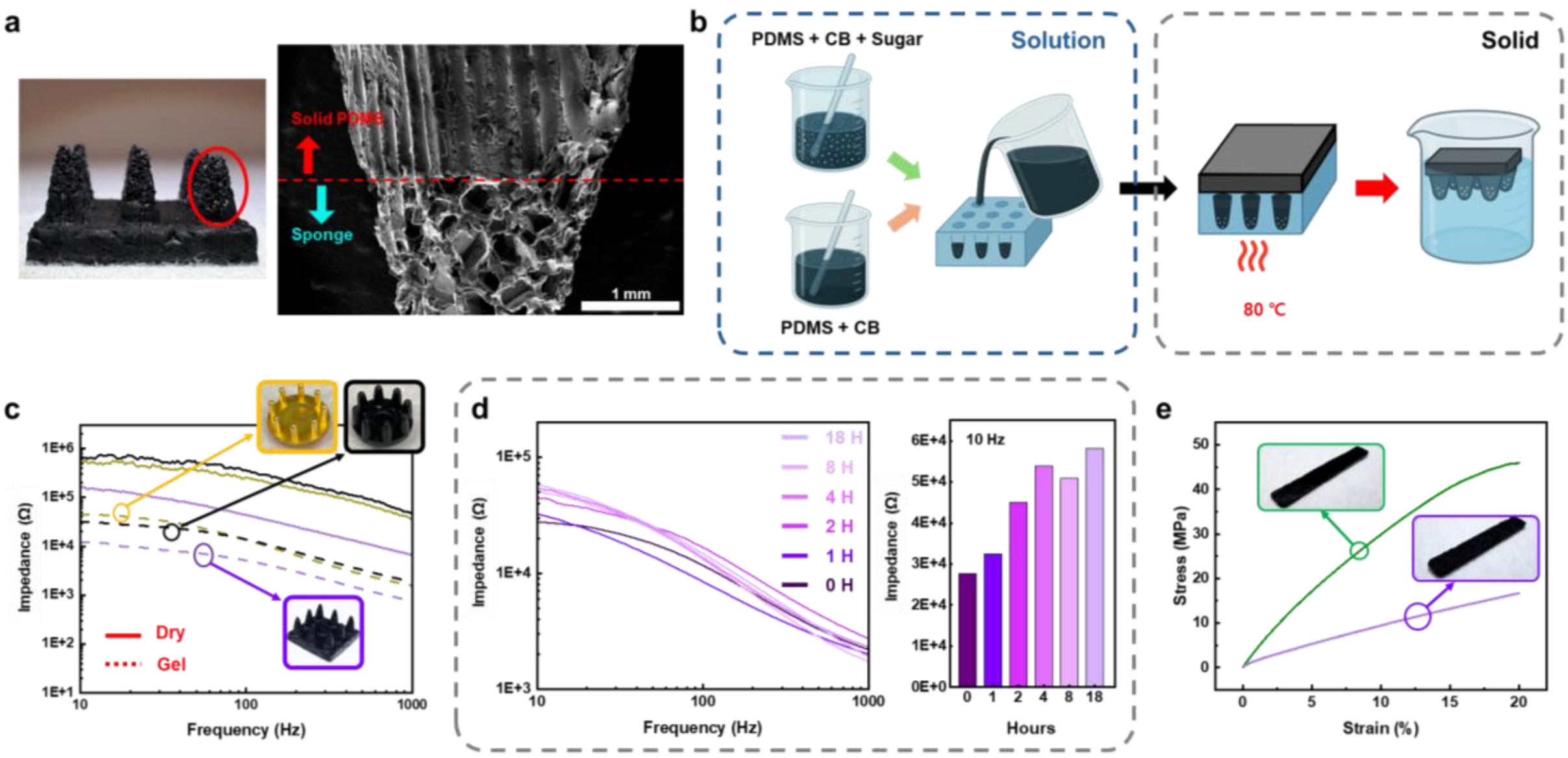

**Figure 2.** Electrical and mechanical properties of the sponge pillar electrode. a) Optical (left) and scanning electron microscope (SEM) (right) images of the two-layered architecture of sponge pillar electrode. b) Schematic of the fabrication process. The sugar-containing precursor is first cast into the mold to form the porous tip layer, followed by the CB-PDMS mixture to form the supporting base. The structure is then thermally cured at 80 °C, and the sugar is subsequently leached out in 80 °C deionized water for 1 hour to generate the sponge like porosity. c) Impedance comparison between sponge pillar, commercial gold, and comb electrodes in both dry and gel-based conditions. The sponge pillar electrode exhibits the lowest contact impedance due to its increased effective contact area. d) Long-term impedance stability of the sponge pillar electrode over 18 hours of ambient exposure. The impedance gradually increases and plateaus after 4 hours, suggesting retained gel hydration. e) Mechanical property comparison between sponge and solid electrodes.

### 2.3. Multi-form-factor EEG signal recording

The flexible and comfortable sponge pillar EEG electrode was evaluated under real-life wearable conditions. The initial frontal EEG recordings were conducted using a commercial OpenBCI Cyton Biosensing Board (**Figure 3a**). As shown in **Figure 3b**, the spectrogram clearly demonstrates the presence of alpha waves from the occipital region (posterior dominant rhythm (PDR)) when the subject closed their eyes. In this recording, the sponge pillar electrode demonstrated comparable performance to the commercial gold electrode placed on the forehead. (**Figure S3**). The presence of alpha waves was also evident in the raw EEG traces (**Figure 3c**), where the red trace, corresponding to the eye closed condition, shows approximately ten cycles of periodic oscillations per second, consistent with the 8-13 Hz alpha wave band. The performance of the sponge pillar electrode was further validated across multiple wearable form factors, including headphones (**Figure 3d**) and a baseball cap (**Figure 3e**). In both cases, the electrode exhibited a distinct alpha band in the spectrograms (Figure S3).

To further improve wearability and reduce noise from wired connections, the electrodes were also tested with a wireless recording system. **Figure 3f** shows the custom designed wireless board used for this evaluation. For compatibility with the wireless system, the sponge pillar electrodes were integrated with commercial Ag/AgCl snap electrodes. The block diagram of the wireless EEG recording system is shown in **Figure 3g**. The analog biopotential signals from the electrodes are first processed through a differential passive filter to suppress DC offsets and high frequency noise. These filtered signals are then amplified by an instrumentation amplifier (TI INA819) with high input impedance and low noise characteristics. The amplified signals are digitized using a 16-bit analog-to-digital converter (TI ADS8681) operating at a 1 kHz sampling rate. A microcontroller (TI CC1310) manages the digitized data, which is then wirelessly transmitted via a sub 1 GHz RF link to a remote receiver. An active shield circuit further enhances signal stability by minimizing external electromagnetic noise interference. This integrated signal chain enables reliable real-time acquisition of high fidelity EEG signals in wearable formats.[33]

The wireless system was also tested using the same form factors as the OpenBCI setup and achieved consistently high SNRs: 12.12 dB for the headband, 12.23 dB for the headphones, and 17.64 dB for the baseball cap (**Figure S4**). To evaluate the electrode's performance under motion conditions, a dynamic recording test was conducted using a headband form factor (**Figure 3h and Video S1**). During the first two minutes, the subject remained seated, keeping their eyes open for one minute and closed for the next minute to induce alpha waves. Afterward, the subject walked back and forth along an 8-meter corridor for another two minutes, with one minute of walking with eyes open followed by one minute of walking with eyes closed to evaluate motion induced artifacts. The sponge pillar electrode maintained high signal quality even during movement. When the subject was seated, the alpha wave SNR was measured at 17.64 dB, and during walking, the SNR only decreased slightly to 16.46 dB. This stability was clearly reflected in the raw EEG traces. As shown in **Figure 3i(i)** (seated) and **Figure 3i(ii)** (walking), the signals exhibited minimal noise oscillations from motion artifacts, demonstrating that the electrode maintained high signal integrity even under ambulatory conditions. In contrast, the commercial comb electrode exhibited substantial noise under motion conditions. As shown in **Figure 3j**, the SNR dropped significantly from 11.64 dB during the stable seated condition to 6.98 dB during walking (**Figure S5**), indicating poor motion tolerance and significant degradation in signal quality. This degradation is more clearly reflected in the raw EEG signals presented in Figure 3k, where the alpha wave signal during the seated condition shows minimal motion induced oscillations (**Figure 3k(i)**), while substantial oscillations caused by motion artifacts are clearly visible when the subject is walking along the corridor (**Figure 3k(ii)**).

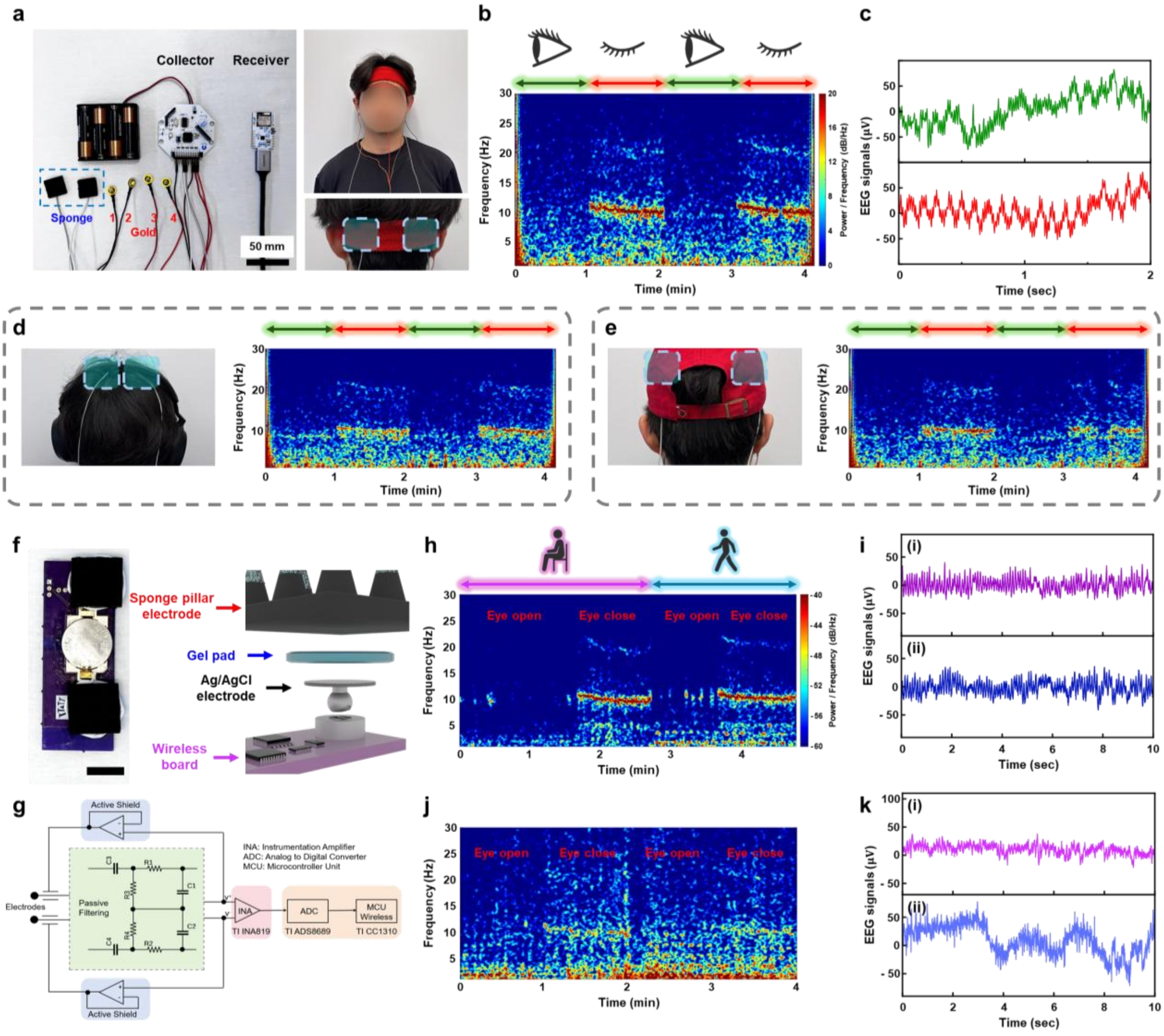

**Figure 3.** EEG recording using sponge pillar electrodes under various wearable form factors and motion conditions. a) Experimental setup for EEG recording using the OpenBCI system. Gold electrodes 1 and 4 were connected to the noise cancelling site (below the left earlobe) and the reference site (right earlobe), respectively. Gold electrodes 2 and 3 were placed on the temple and forehead to enable cross validation of EEG. b) Time-frequency spectrogram of EEG signals measured using a headband integrated sponge pillar electrode during alternating eye open and close states. c) Corresponding raw EEG signals from the headband

measurement, showing clear alpha wave emergence (8-12 Hz) during eyes closed periods (red) compared to eyes open states (green). d, e) Time-frequency spectrograms of EEG signals measured using the same protocol as in b), but with the sponge electrode integrated into headphones d) and a baseball cap e), respectively. d) and e) share the same color bar of b). f) Illustration of wireless EEG system integrating the sponge pillar electrode, hydrogel pad, Ag/AgCl reference electrode, and a custom designed low noise wireless board. Scale bar, 10mm. g) Schematic diagram of the analog signal processing and data transmission pathway. h) Spectrogram of EEG recorded with the wireless system while transitioning between eyes open and eyes closed states during both resting (purple) and walking (blue) conditions. i) Raw EEG signals from channel (i) and (ii) corresponding to the eyes closed segments in Figure 3h). Panel (i) represents the seated condition, while panel (ii) shows EEG recorded during walking. j) Spectrogram of EEG signals recorded using commercial electrodes during repeated eye open/close cycles while walking. k) Corresponding raw EEG signals from (i) seated and (ii) walking conditions during eyes closed states shown in Figure 3j) and Figure S5. (ii) demonstrates substantial motion artifacts, highlighting the limitations of conventional electrodes under dynamic conditions. For ambulatory EEG recordings, raw signals without additional smoothing and filtering are presented to facilitate direct comparison of motion induced artifacts.

## 2.4. Multi-channel EEG mapping

To extend the sponge pillar electrode from a single channel configuration to a multichannel EEG system, 19 sponge pillar electrodes were integrated with a flexible, screen-printed textile electrode patch. The assembled EEG patch was connected to a passive BioSemi system via a custom designed flexible printed circuit board (FPCB) connector and mounted on the central crown region inside a baseball cap using double sided adhesive tape (Figure 1b, **Figure 4a, and 4b**). During the recording session, the subject alternated between one minute of eyes open and one minute of eyes closed, repeatedly over a 10-minute period, to elicit alpha wave modulation. One active BioSemi electrode simultaneously placed adjacent to the sponge pillar electrode for comparison with commercial electrode. As shown in **Figure S6**, the passive sponge electrode demonstrated comparable alpha wave responses throughout the 10-minute test. In addition, all 19 channels successfully recorded distinct alpha activity during the eye closed periods (**Figure S7**). The recorded EEG signals were processed to generate spatial mapping of alpha (8-13 Hz) band power, focusing on the transitions between cognitive states. The absolute power in decibels was calculated within each sliding window without global normalization. This dynamic mapping clearly visualizes how band power evolves during the transitions between eyes open and eyes closed state. **Figure 4c(i)** shows a gradual reduction of alpha power as the subject transitions from eye closed to eye open, while **Figure 4c(ii)** illustrates a progressive increase in alpha activity during the reverse transition. These spatiotemporal dynamics are further demonstrated in **Video S2**. These results highlight the potential of the sponge pillar electrode for wearable multichannel EEG mapping, enabling reliable tracking of cortical state transitions under practical conditions.

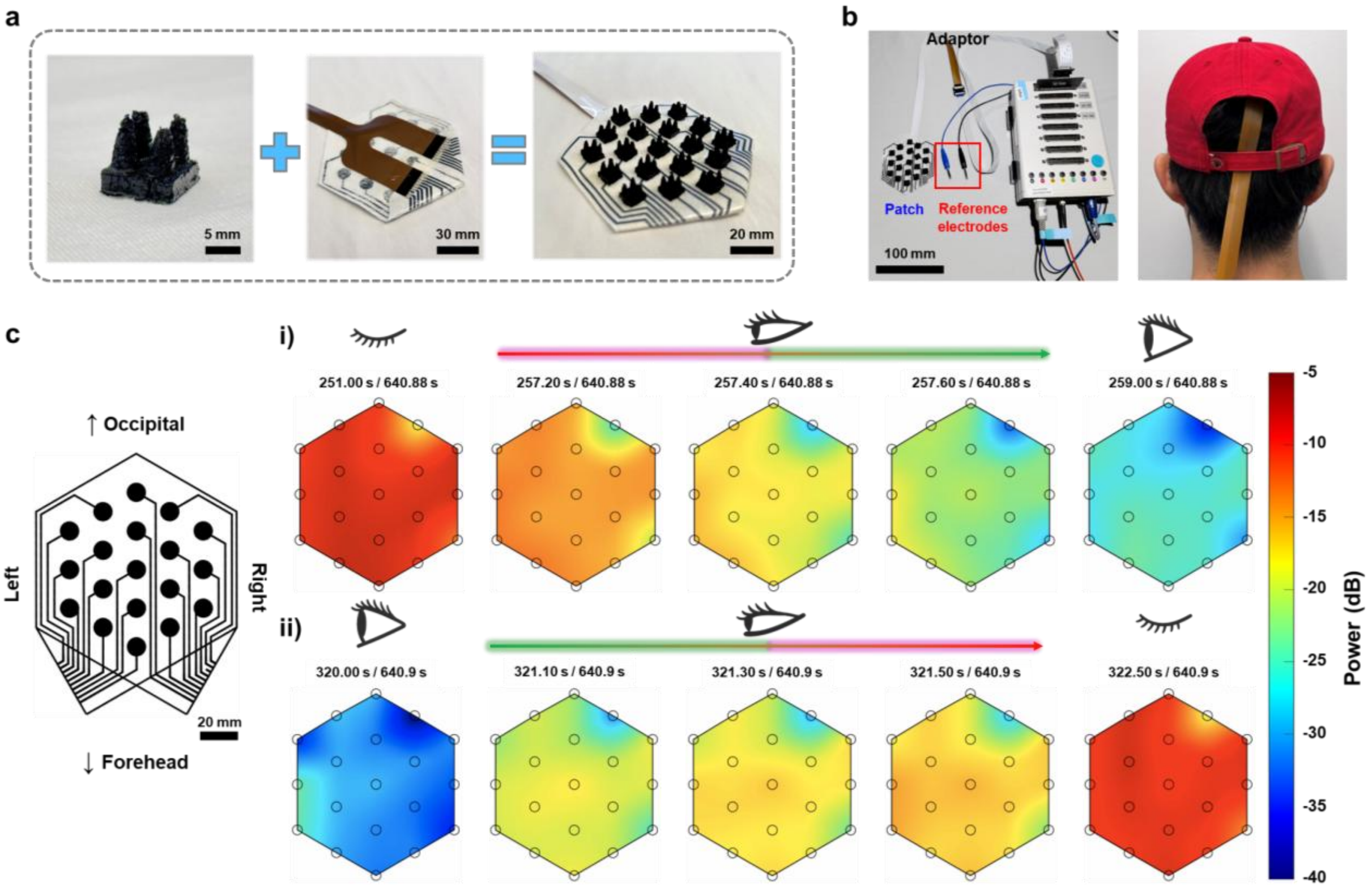

**Figure 4.** Multichannel EEG mapping using sponge pillar and textile hybrid electrodes. a) Photograph of the integration of sponge pillar electrodes with a screen-printed PEDOT:PSS textile patch for multichannel configuration. b) Image of the completed 19 channel electrode array connected to the BioSemi system via a custom-designed adaptor, implemented in a wearable EEG setup using a baseball cap. The reference electrodes were placed below the left and right earlobes. c) Layout of the 19-channel textile electrode array. Time resolved spatial maps of alpha band activity during repeated transitions: (i) eyes closed to open and (ii) eyes open to closed.

## 2.5. Sleep analysis

The recording performance of the sponge pillar electrode was further evaluated in sleep EEG measurements, which require the reliable acquisition of both high frequency (> 10 Hz) and low frequency (< 5 Hz) signals. Sleep EEG recordings were conducted using all three wearable form factors (headband, headphones, and baseball cap). Our custom-designed wireless recording system was used for the headphones (**Figure S8**) and baseball cap form factors, whereas the OpenBCI wired system was employed for the headband form factor, as it allowed for electrode placement on hairy scalp regions (**Figure S9**). **Figure 5a** shows the subject in a sleeping position with the electrode placement. At the beginning of the session, the subject lay on a recliner with eyes closed in a light controlled environment with the lights turned off, and the study was performed in the morning after partial sleep restriction on the preceding night. **Figure 5b** shows the spectrogram of the recorded EEG over a total duration of 179 min. Based on this recording, sleep staging was performed to generate the hypnogram shown in **Figure 5c**, while sleep stage information was simultaneously acquired using an Apple Watch, as shown in **Figure 5d**. Because the EEG was recorded primarily from crown area, canonical rapid eye movement (REM) markers were difficult to distinguish reliably from the recording alone, and the analysis therefore focused on non-rapid eye movement (NREM) associated EEG features. During the first 9 minutes of the recording, the subject remained awake, as evidenced by the presence of a strong PDR visible in the spectrogram. Following this awake period, the PDR gradually diminished and became more dispersed, indicating the subject was transitioning into sleep. After sleep onset, the spectrogram clearly exhibited characteristic EEG features of N2 sleep, including sleep spindles (12-15 Hz), followed by low frequency activity characteristic of N3 sleep, including delta waves (1-4 Hz).[34] A transition from deep NREM sleep to REM sleep occurred approximately one hour after sleep onset, characterized by a reduction in delta power and a noticeable emergence of theta oscillations (4-7 Hz), which are typical of REM sleep. Subsequently, the participant transitioned back into a deeper sleep stage, as reflected by an increase in low frequency EEG activity. Around 2 hours and 10 minutes after sleep onset, the Apple Watch recorded a brief awake period lasting approximately 2 minutes. During this interval, the spectrogram did not show a clear reappearance of PDR typically associated with wakefulness but instead showed a transient reduction in low frequency power. Consistent with this, hypnogram suggested a brief fluctuation between N3 and N2 around the same period, although the transition was not sharply defined. Following this brief awakening, the subject re-entered deep sleep and subsequently progressed into lighter sleep stages, as evidenced by dynamic fluctuations in delta power over time. Representative raw EEG traces showing PDR during wakefulness, sleep spindles during N2 sleep, and delta waves during N3 sleep are provided in **Figure S10**. Notably, no obvious signs of sleep fragmentation suggestive of discomfort were observed during the recording, indicating that wearing sponge pillar electrodes did not noticeably disrupt sleep continuity.

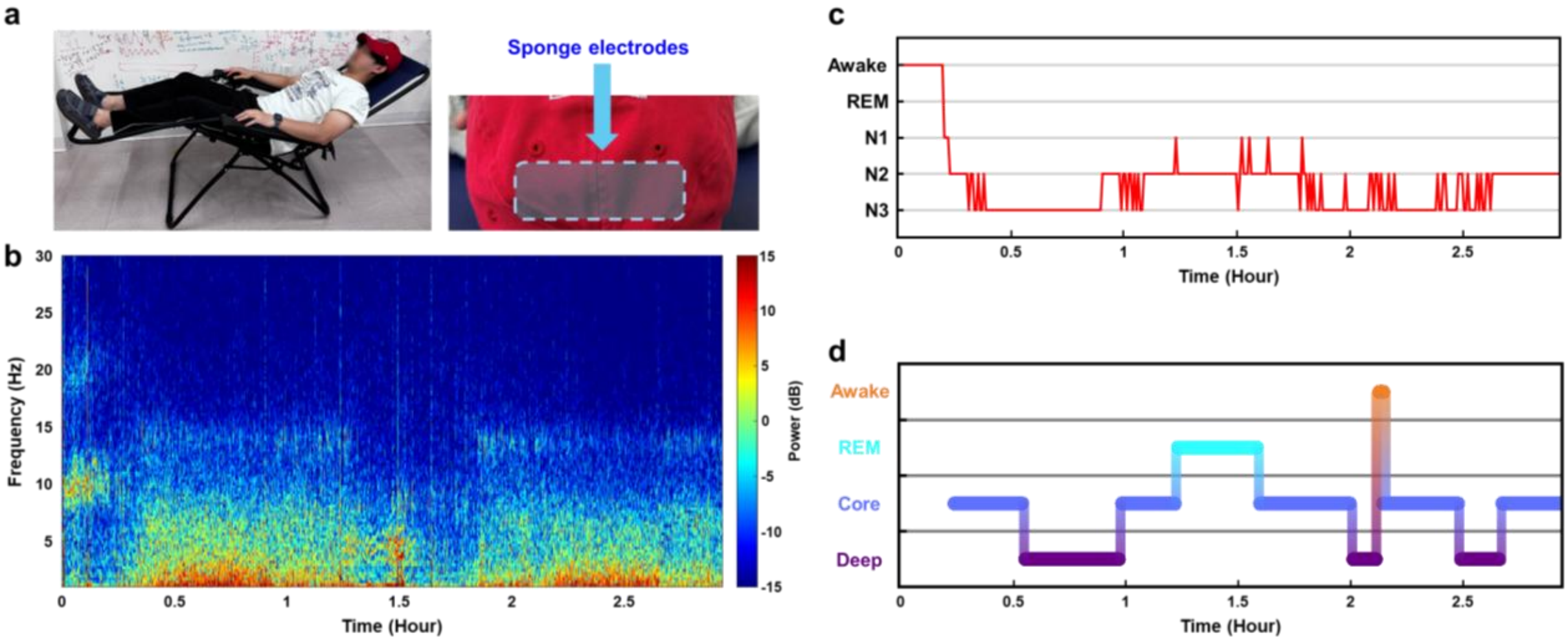

**Figure 5.** Wireless sleep EEG recording and analysis. a) Photograph of the subject during the sleep EEG recording and the corresponding electrode placement. b) Spectrogram of the recorded EEG over the full recording period. c) Hypnogram showing the sleep stage transition estimated from the EEG recording. d) Sleep stage information obtained simultaneously from the Apple Watch.

## 3. Conclusion

This study presents a significant advancement in wearable EEG technology through the development of sponge pillar electrodes composed of CB and PDMS, integrated with PEDOT:PSS screen-printed textile substrates. The unique two-layer configuration featuring a soft microporous sponge tip and a firmer base layer enables effective penetration through hair dense regions and enhances the skin-electrode contact area. This design achieves reduced contact impedance under gel conditions and demonstrates excellent gel retention in ambient air, maintaining signal stability over extended durations. Furthermore, the mechanical softness of the sponge structure minimizes motion artifacts while improving wearer comfort. The electrode was adapted to various practical and unobtrusive wearable form factors including headbands, headphones, and baseball caps and successfully recorded EEG signals from different scalp regions, such as the occipital, vertex, and temporal areas, even under ambulatory conditions. The textile electrode platform not only facilitated seamless integration into wearable designs but also enabled the construction of a 19-channel EEG system for spatiotemporal mapping of brain activities. Finally, the electrode demonstrated reliable performance during sleep recordings, maintaining user comfort and delivering high quality signals over prolonged periods. This innovative EEG electrode system paves the way for the next generation of wearable EEG systems, providing a wearable, high performance, and user-friendly platform for real-time monitoring in neuroscience, clinical diagnostics, and neurotechnology applications.

## 4. Experimental Section

### 4.1. Materials

PEDOT:PSS aqueous dispersion (5 wt%) was obtained from Sigma Aldrich (St. Louis, MO). Silver epoxy adhesive (8331D) was purchased from MG Chemicals (Ontario, Canada). Sylgard 184 PDMS elastomer kit was acquired from Dow Corning (Midland, MI). Conductive carbon black (Super P) was sourced from MSE Supplies (Tucson, AZ). Commercial gel pads were provided by Hollywog (Chattanooga, TN), and Ag/AgCl reference electrodes were obtained from 3M (St. Paul, MN). Cleanroom textile substrate (TX 609) was sourced from TechniCloth (Kernersville, NC). White sugar used as a sacrificial template was procured from Domino Foods (Yonkers, NY).

### 4.2 Instruments

Surface morphology was characterized using an environmental scanning electron microscope (Quattro S ESEM, Thermo Fisher, UK). Screen-printed electrodes were fabricated using a semi-automatic screen printer (DP-320, Newlong, Japan). Electrode resistance was measured using a digital multimeter (Fluke 117, Fluke Corp., Everett, WA). Skin-electrode impedance was recorded using an impedance analyzer (Bode 100, OMICRON Lab, Austria). Customized molds were 3D printed using a Prusa MK4 system (Prague, Czech Republic). Mechanical characterization was performed using a low force testing system (Electroforce 3200, TA Instruments, New Castle, DE). EEG signals were acquired using either the Cyton Biosensing Board (OpenBCI, Brooklyn, NY), BioSemi system (Amsterdam, Netherlands), or an in-house developed wireless system (Chakrabartty Lab, Washington University in St. Louis, MO).

### 4.3 Electrode fabrication

*Conductive elastomer paste.* To construct conductive elastomer for the base and substrate layers, PDMS base and curing agent were mixed at a 10:1 weight ratio and stirred for 5 minutes. Carbon black was then added incrementally to a concentration of 12.5 wt% and further mixed for 5 minutes at room temperature to yield a uniform conductive elastomer composite.

*Conductive sponge elastomer paste.* To fabricate the conductive sponge elastomer, powdered white sugar was added at a 1:3.5 weight ratio and hand mixed for 5 minutes to produce a thick paste. This composite served as the porous lower structure of the sponge pillar.

*Sponge pillar electrode.* A two-layer pillar electrode was fabricated in a custom designed cylindrical plastic mold. It was filled halfway with conductive sponge paste and compressed using a 2 mm diameter rod to achieve consistent pillar geometry. The remaining portion was filled with the conductive elastomer paste to form the base. The full structure was thermally cured at 50 °C for 12 hours and post cured at 80 °C for 1 hour. The sugar template was then removed by immersing the structure in deionized water maintained at 80 °C for 1 hour, after which the electrode was thoroughly dried in a convection oven at 80 °C to eliminate residual moisture.

*Textile electrode preparation*. Textile electrodes were fabricated by screen printing PEDOT:PSS onto pretreated textile substrates through a micromesh mask, followed by thermal annealing at 80 °C for 1 hour to promote film uniformity and adhesion. Silver epoxy was applied to form contact pads, and flexible flat cables were connected for integration into EEG acquisition systems.

*Integration of textile and sponge pillar electrodes*: For multi-site EEG recording, 19 sponge pillar electrodes were integrated onto the screen-printed textile electrodes using a soft ionic gel pad to form a conformal assembly.

### 4.4 Characterization of Electrodes

*Electrode-skin contact impedance characterization.* To evaluate electrode-skin interface impedance, a pair of electrodes was placed on the volar forearm surface with an interelectrode spacing of 8 cm. Impedance spectra were recorded across a frequency range from 1 Hz to $10^4$ Hz.

*Mechanical properties of electrode.* To evaluate the mechanical properties, both sponge and solid CB-PDMS electrodes were fabricated with identical dimensions (width: 8 mm, length: 50 mm, thickness: 2 mm). Each specimen was fixed at one end using a mechanical mount, while the opposite end was attached to a 40 N load cell. Uniaxial tensile testing was conducted with a maximum extension of 13 mm at a constant elongation rate of 0.083 mm/s.

### 4.5 EEG signals recording

The EEG recording was conducted at the Washington University in St. Louis, Dr. Shantanu Chakraborty's laboratory (IRB #: 201907109).

*OpenBCI EEG signal recording setup.* For EEG signal acquisition, two electrodes were consistently positioned as follows:

1) Reference Electrode: Placed on the right earlobe.
2) Noise Canceling Electrode: Positioned below the left earlobe.

Simultaneously, two gold electrodes were used as a comparison standard to cross check the performance of the sponge pillar electrode.

*Wireless board EEG signal recording system.* sponge pillar electrodes were adhered to commercial Ag/AgCl electrodes (with adhesive rim removed). These assemblies were connected to a custom wireless board designed for EEG data acquisition, eliminating the need for separate reference electrodes.

*Head band EEG form factor.* For the OpenBCI system, a total of four electrodes were secured beneath a fabric headband using 3M double sided tape. One gold electrode was placed on the forehead and the other on the left temple. Two sponge pillar electrodes were symmetrically positioned 3 cm lateral to the occipital midline on the posterior scalp. For recordings using the wireless board system, the board was positioned at the occipital center.

*Headphones EEG form factor.* In the OpenBCI based setup, two gold electrodes were affixed to the forehead and temple beneath the head pad, and two sponge pillar electrodes were positioned at the parietal region, symmetrically 3 cm from the vertex. For the wireless board system, the recording board was aligned to the vertex (top of the scalp), with the sponge electrodes.

*Baseball cap EEG form factor.* For real-world application testing, electrodes were embedded beneath the crown region of a standard baseball cap. In the OpenBCI setup, electrode positions mirrored those in the headband configuration, with forehead and temple placement for gold electrodes and mid-occipital positioning for sponge electrodes.

For wireless board integration, the cap was modified to house the board at the vertex, with sponge pillar electrodes secured on either side of the center using double sided adhesive.

*Multichannel EEG measurement.* For multichannel EEG acquisition, a textile electrode patch containing multiple recording sites was integrated inside the crown region of a standard baseball cap. The flexible flat cable extending from the patch was connected to a custom designed adapter, which interfaced with a passive BioSemi system. A 10-minute recording session was conducted while the subject remained seated and relaxed.

*Sleep EEG measurement.* Sleep EEG recordings were performed while the subject rested in a reclined chair. Sponge pillar electrodes were placed at the same positions as in the headband (for OpenBCI), headphones, and baseball cap (for wireless board) configurations. For the headphones and cap setups, an Apple Watch was used simultaneously to monitor sleep for cross checking purposes.

**4.6 EEG signals processing**

*OpenBCI EEG signal processing.* All OpenBCI EEG data were sampled at 250 Hz, and the corresponding time vector was generated accordingly. The raw signals were preprocessed using a bandpass filter between 0.5 and 100 Hz to remove baseline drift and high frequency noise. To analyze EEG signals, spectrograms were computed for each electrode using a 1000-point Hamming window (equivalent to 4 seconds at 250 Hz sampling rate) with 90% overlap. The analysis focused on the 0.5-30 Hz range to highlight key physiological rhythms during prolonged recordings. Additionally, for sleep EEG analysis, the log magnitude spectrum was computed using a base 10 logarithm

$$10 \cdot \log_{10}(|\mathrm{FFT}(x)| + \varepsilon)$$

where $\varepsilon$ is a small constant added to prevent singularity at zero.

*Wireless EEG signal processing.* Wireless EEG recordings were sampled at 1000 Hz, and the time vector was constructed accordingly. A bandpass filter (0.5-100 Hz) was applied, followed by spectrogram analysis using a 4096-point Hamming window with 90% overlap. The resulting spectrograms were visualized over the 0.5-30 Hz frequency range. These spectral representations were analyzed under both stationary and walking conditions to compare EEG signal quality. For visualization purposes (Video S1), unfiltered raw EEG signals were used instead of bandpass filtered data. Sleep EEG analysis was performed using the same log magnitude spectral approach as described in the OpenBCI section.

*Passive BioSemi EEG signal processing.* EEG signals recorded using the BioSemi system were sampled at an original rate of 2048 Hz. To extract EEG components within the 0.5-50 Hz range, the signals were downsampled to 128 Hz following a sliding average, with the window size determined by the ratio of the original sampling rate to the target rate. The spectrograms were computed using the pspectrum function in MATLAB. Each spectrogram was generated with a 90% overlap and a frequency resolution of 1 Hz, and was visualized over the 0.5-30 Hz range.
Alpha (8-12 Hz) band power distributions were computed and visualized over time using electrode level EEG signals. The resulting power spectral density (PSD) was converted to decibel scale using the same base 10 logarithmic transformation as described for the OpenBCI system. Band specific power was computed within 1 second sliding windows (step size: 10 ms) by averaging the PSD values over the respective frequency ranges and time bins. The resulting spatiotemporal maps were compiled into time lapse videos rendered at 100 frames per second (Video S2).

*Statistical analysis of the SNR.* To quantitatively assess the signal quality of alpha band across EEG channels, a spectrogram-based signal-to-noise ratio (SNR) was calculated using time-frequency decomposition. Prior to spectral estimation, each EEG signal was z-score normalized to ensure zero-mean and unit variance. For each channel, the alpha band signal power was estimated by averaging the power spectral density values within 8-12 Hz across a 60-second analysis window.[35] To estimate the background noise, power values from a broader range (5-35 Hz) were used while excluding the signal band (7-13 Hz) and its harmonics (18-22 Hz) to avoid contamination. The SNR for each channel was then computed as the difference between the time averaged dB scaled signal and noise powers:

$$SNR_{dB} = \overline{P}_{alpha\ (dB)} - \overline{P}_{noise\ (dB)}$$

Where $\overline{P}_{alpha\ (dB)} = \frac{1}{T}\sum_{t=1}^{T} 10log_{10}(P_\alpha(t) + \varepsilon)$ and similarly for $\overline{P}_{noise\ (dB)}$. Here, $\overline{P}_{alpha\ (dB)}$ is the instantaneous power in the alpha band at time t, and ε is a small positive constant to avoid numerical instability. T denotes the total number of time samples within the analysis window, calculated as the product of the sampling frequency and the duration in seconds ( $T = f_s \times 60$ for a 60 second window).

*Sleep scoring.* The data were downsampled to 125 Hz and converted to EDF format. The recordings were filtered from 0.3 to 35 Hz after being imported into Philips Sleepware G3 software. They were scored in 30-second epochs by registered polysomnographic technologists using American Academy of Sleep Medicine (AASM) guidelines.

## Supporting Information

Supporting Information is provided at the end of this manuscript.

## Acknowledgements

This work was funded by the NIH/Eunice Kennedy Shriver National Institute of Child Health and Human Development under award number R01HD105905 and NIH/National Institute of Biomedical Imaging and Bioengineering under award number R01EB038720. The authors acknowledge the Institute of Materials Science and Engineering at Washington University for the use of instruments and staff assistance.

## Conflict of Interest

The authors declare no conflict of interest.

## Data Availability Statement

All data supporting the results in this study are provided in the paper and Supplementary Information. Additional data are available from the corresponding authors upon reasonable request.

# Supporting Information

**A Soft Sponge-Pillar Electrode Platform for Motion-Tolerant Ambulatory EEG and Sleep Monitoring**

*Chansoo Kim[†], Weilun Li[†], Ryan Andersen, Yiheng Zhang, Shuoyan Liu, Junyi Zhao, Yuelin Li, Zhongyu Luo, Josephine Lau, Yang Huang, Diane Bracken, Matthew Singh, ShiNung Ching, Ben Julian A. Palanca, Yong Wang, Shantanu Chakrabartty, and Chuan Wang**

Chansoo Kim, Weilun Li, Ryan Andersen, Yiheng Zhang, Shuoyan Liu, Junyi Zhao, Yuelin Li, Matthew Singh, ShiNung Ching, Shantanu Chakrabartty, and Chuan Wang
Electrical & System Engineering, Washington University in St. Louis, St. Louis, Missouri, United States of America
E-mail: chuanwang@wustl.edu

Chansoo Kim and Chuan Wang
Institute of Materials Science and Engineering, Washington University in St. Louis, St. Louis, Missouri, United States of America

Zhongyu Luo
Image Science, Washington University in St. Louis, St. Louis, Missouri, United States of America

Josephine Lau
Biomedical Engineering, Washington University in St. Louis, St. Louis, Missouri, United States of America

Diane Bracken and Ben Julian A. Palanca
Anesthesiology, Washington University School of Medicine, St. Louis, Missouri, United States of America

Zhongyu Luo, Josephine Lau, Yuelin Li, Yang Huang, and Yong Wang
Obstetrics and Gynecology, Washington University School of Medicine, St. Louis, Missouri, United States of America

† Chansoo Kim and Weilun Li contributed equally to this work.

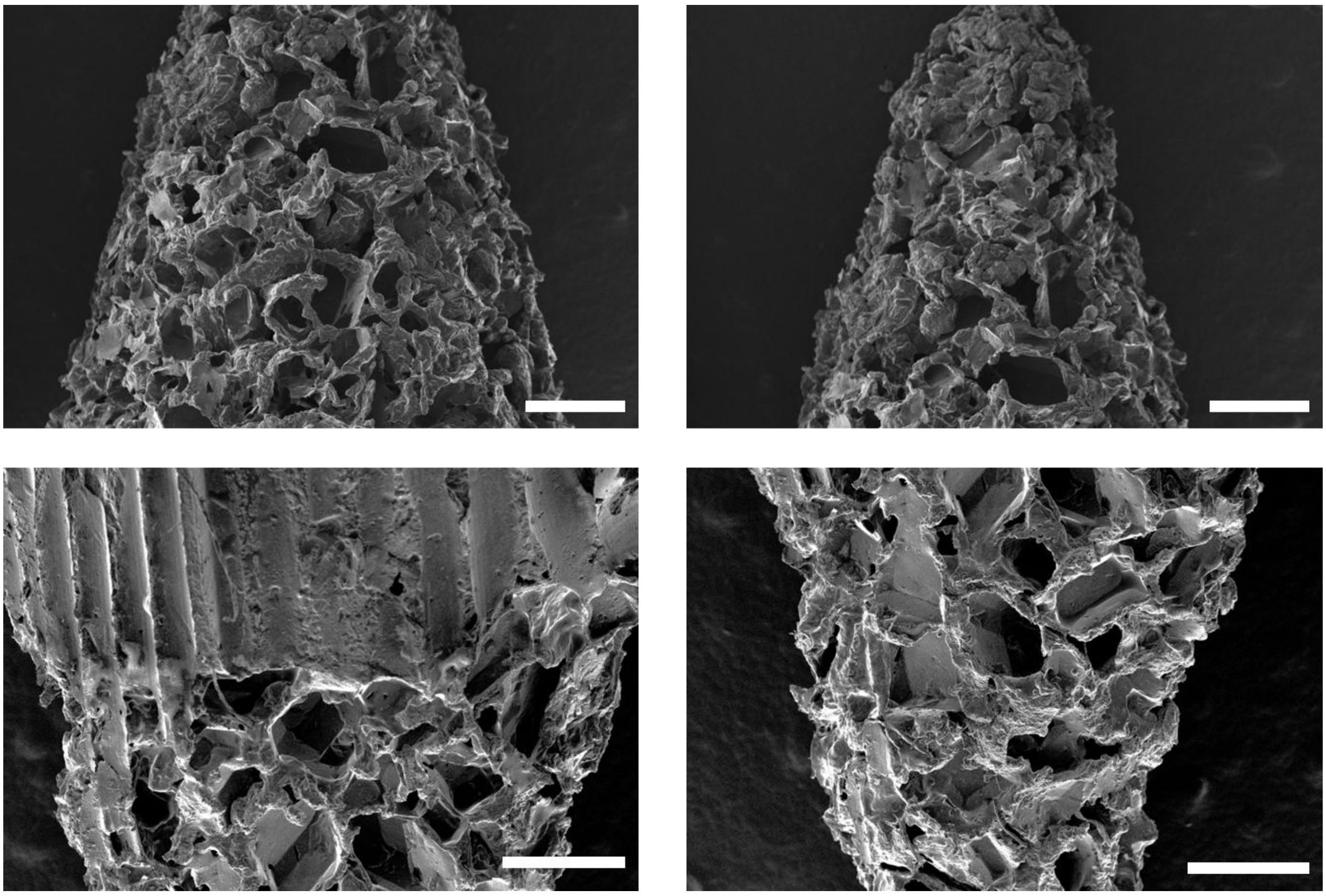

**Figure S1.** Scanning electron microscope (SEM) image of sponge pillar electrode. The average pore size of sponge tip is 258 μm. Scale bars, 500μm

To evaluate the long-term gel retention capability of the sponge pillar electrode, which maintains internal moisture by stably storing conductive gel within its porous structure, EEG signals were recorded in parallel with the measurements shown in Figure 2d.[1] After a defined period of exposure to ambient conditions, each electrode was placed on the occipital region of the subject using a headband following impedance characterization.

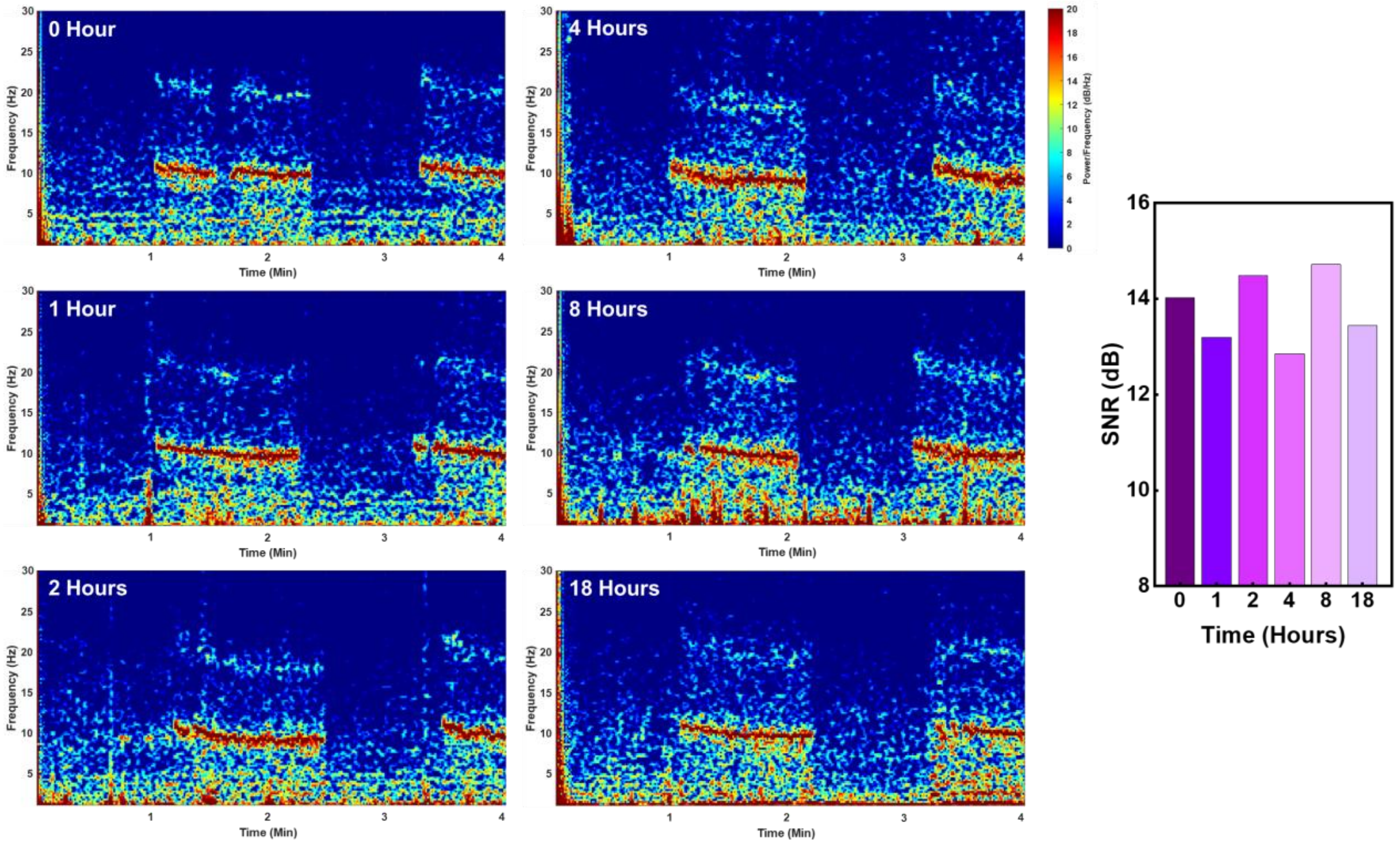

**Figure S2.** Long-term stability of sponge pillar electrode. The subject repeatedly opened and closed their eyes at 1-minute intervals to induce alpha wave activity. SNRs were calculated based on the 1-minute eye closed segments.

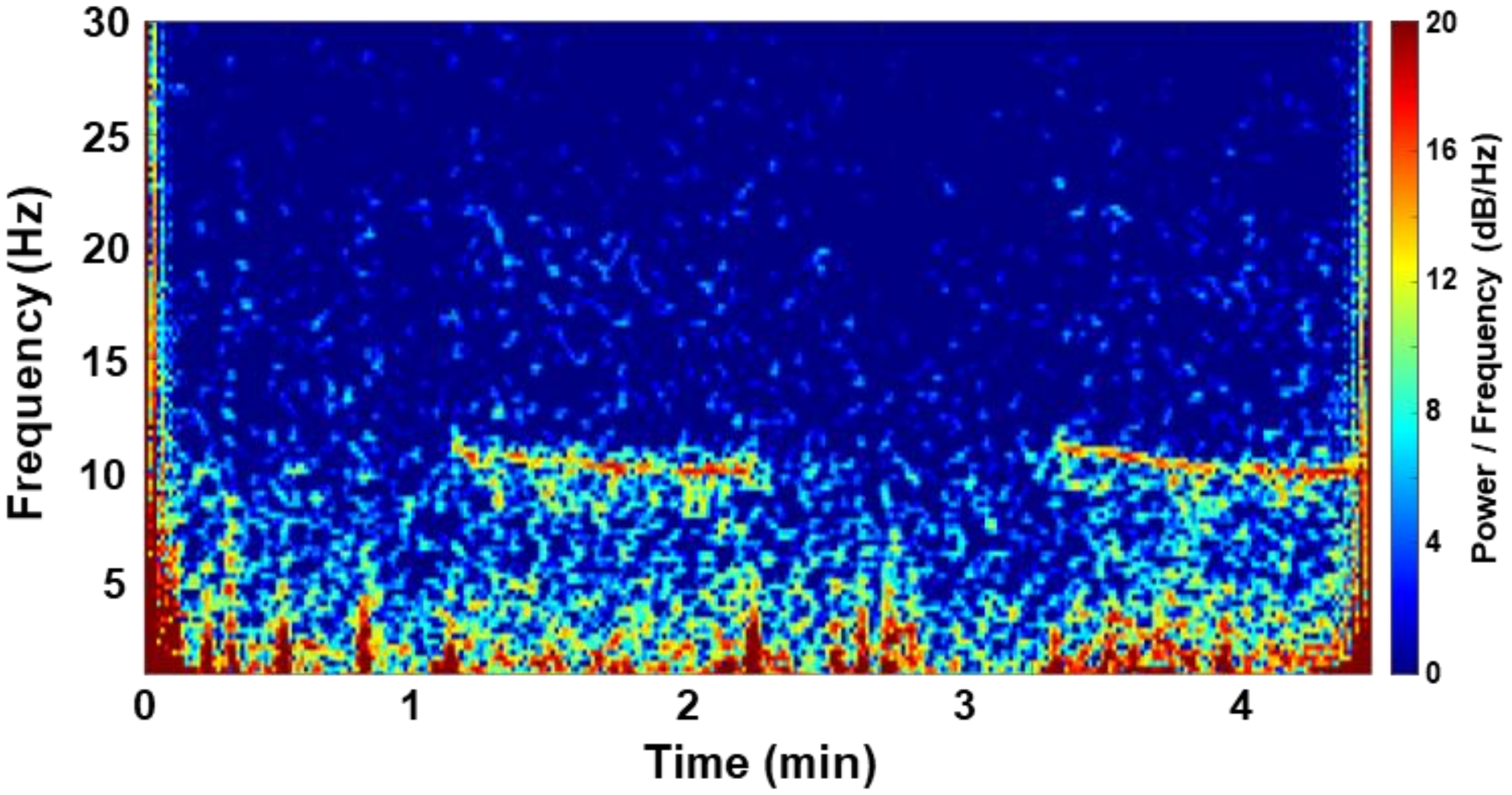

**Figure S3.** EEG alpha wave recording using commercial gold cup electrodes. Two additional gold cup electrodes were simultaneously placed on the subject's forehead and temple during the sponge pillar electrode measurements for cross validation purposes.

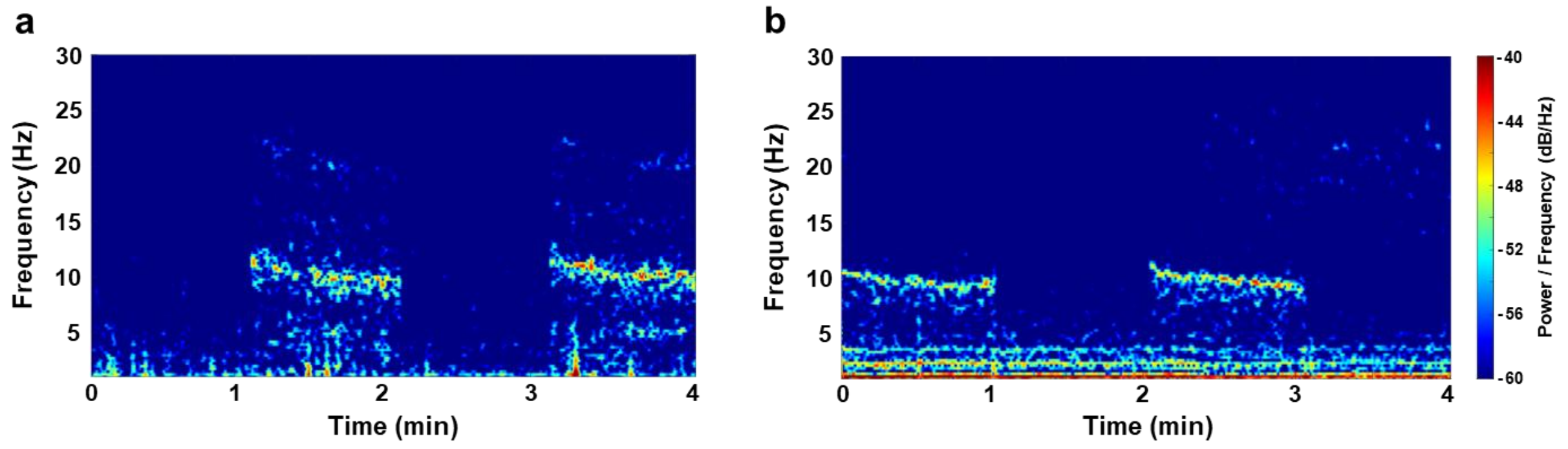

**Figure S4.** Alpha wave recordings acquired from the sponge pillar electrode using a wireless board. a) Headband form factor. b) Headphones form factor. The two spectrograms share the same color bar.

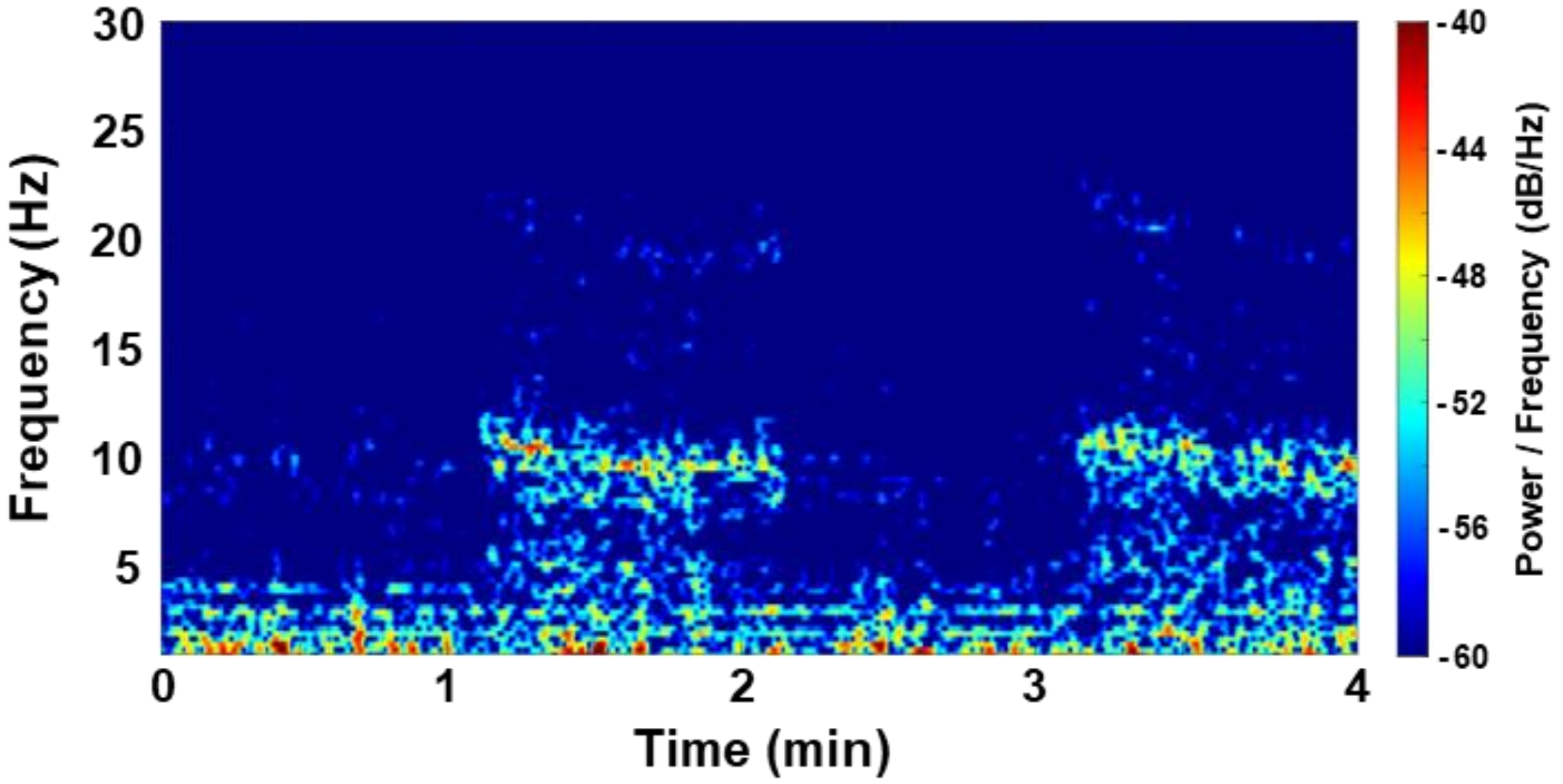

**Figure S5.** EEG signal from a commercial comb electrode. Recording was conducted with the subject seated. The electrode was connected to an Ag/AgCl snap electrode via gel pad, as in Figure 3f.

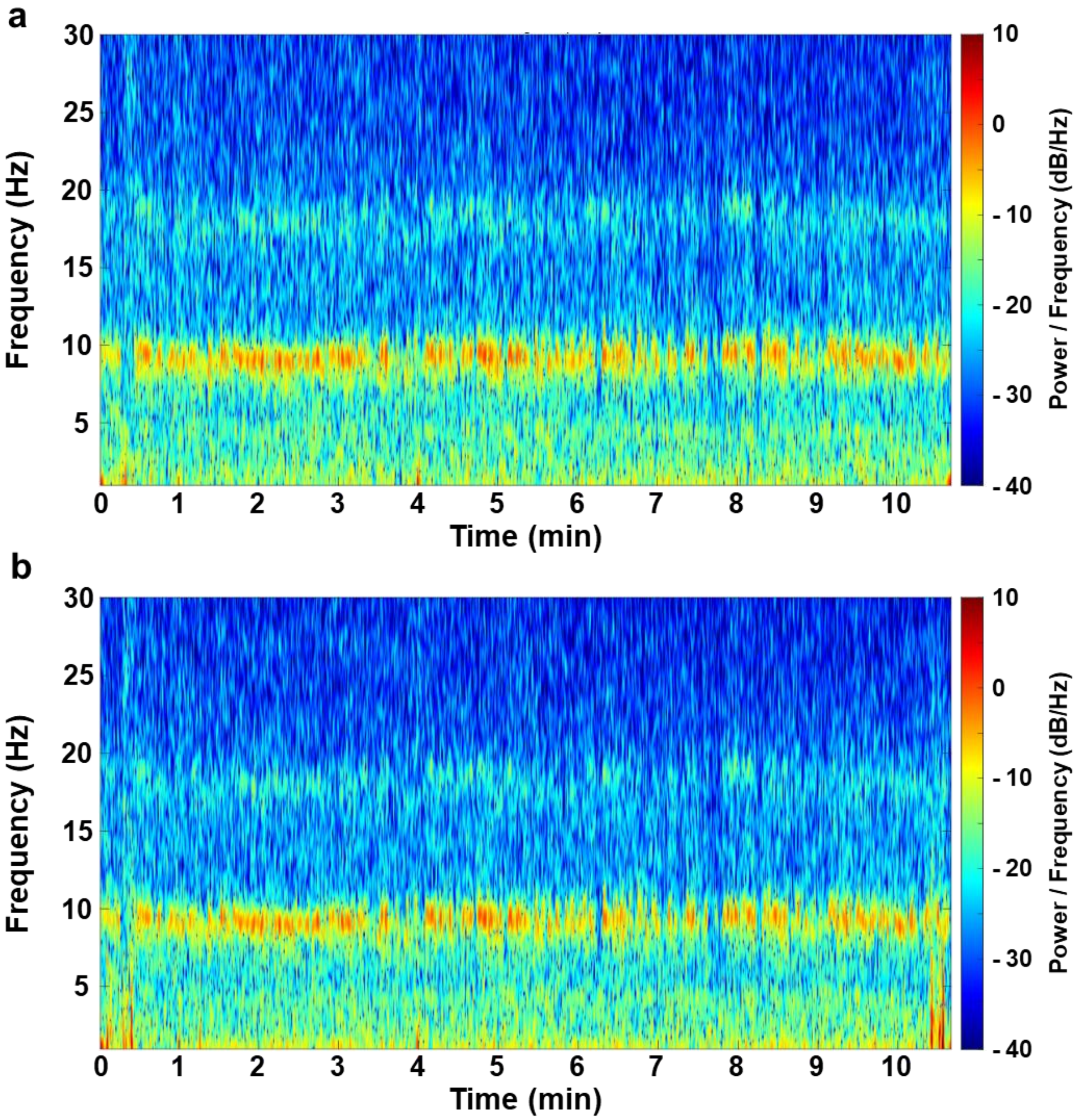

**Figure S6.** EEG signal comparison. a) Commercial BioSemi active EEG electrode. b) System integrated with a sponge pillar and textile electrode patch. The sponge electrode patch exhibited comparable signal quality to the commercial active electrode.

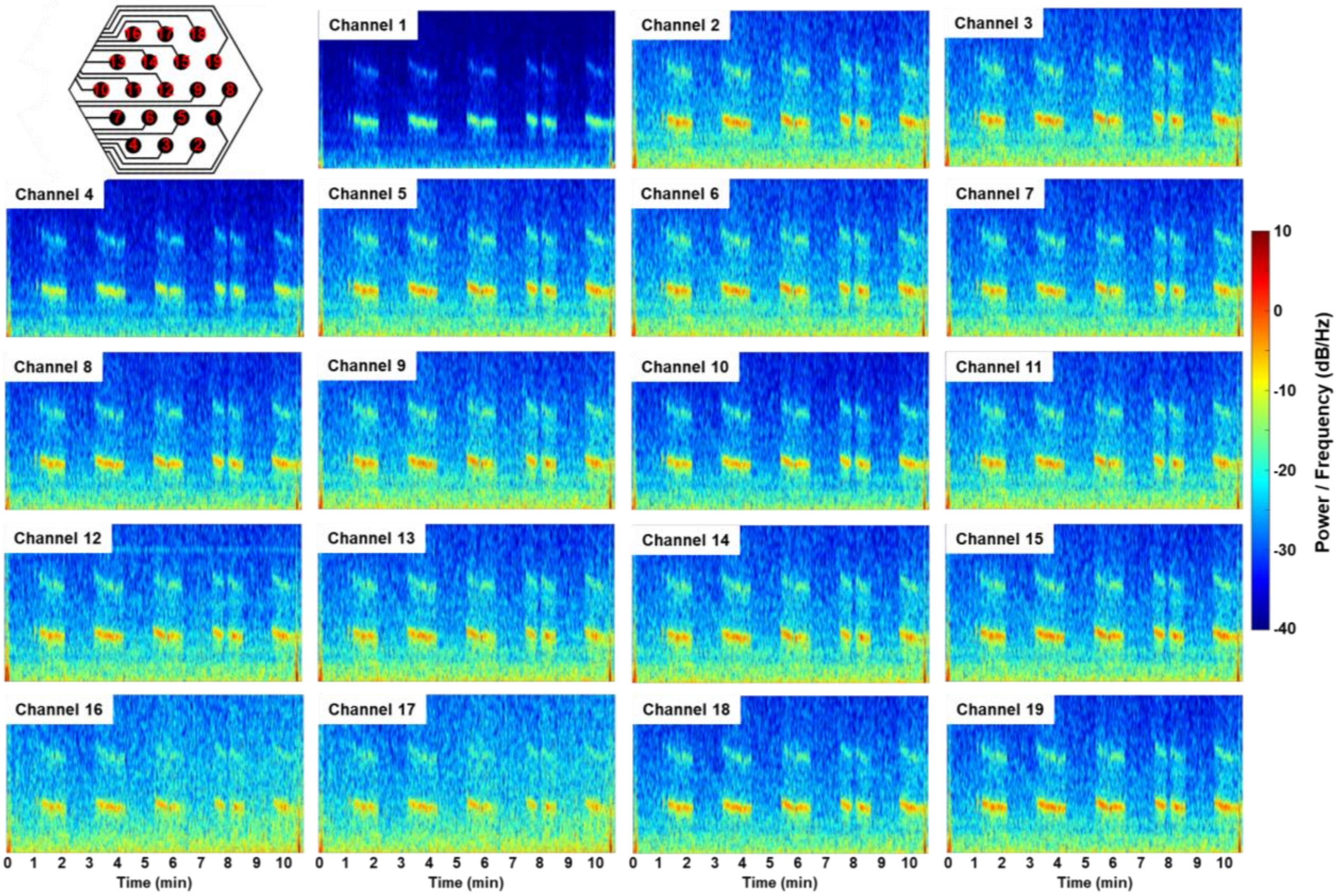

**Figure S7.** Multichannel EEG recording using a passive BioSemi system. EEG signals were successfully recorded from all 19 channels during a 10-minute session. All channels share the color bar on the right.

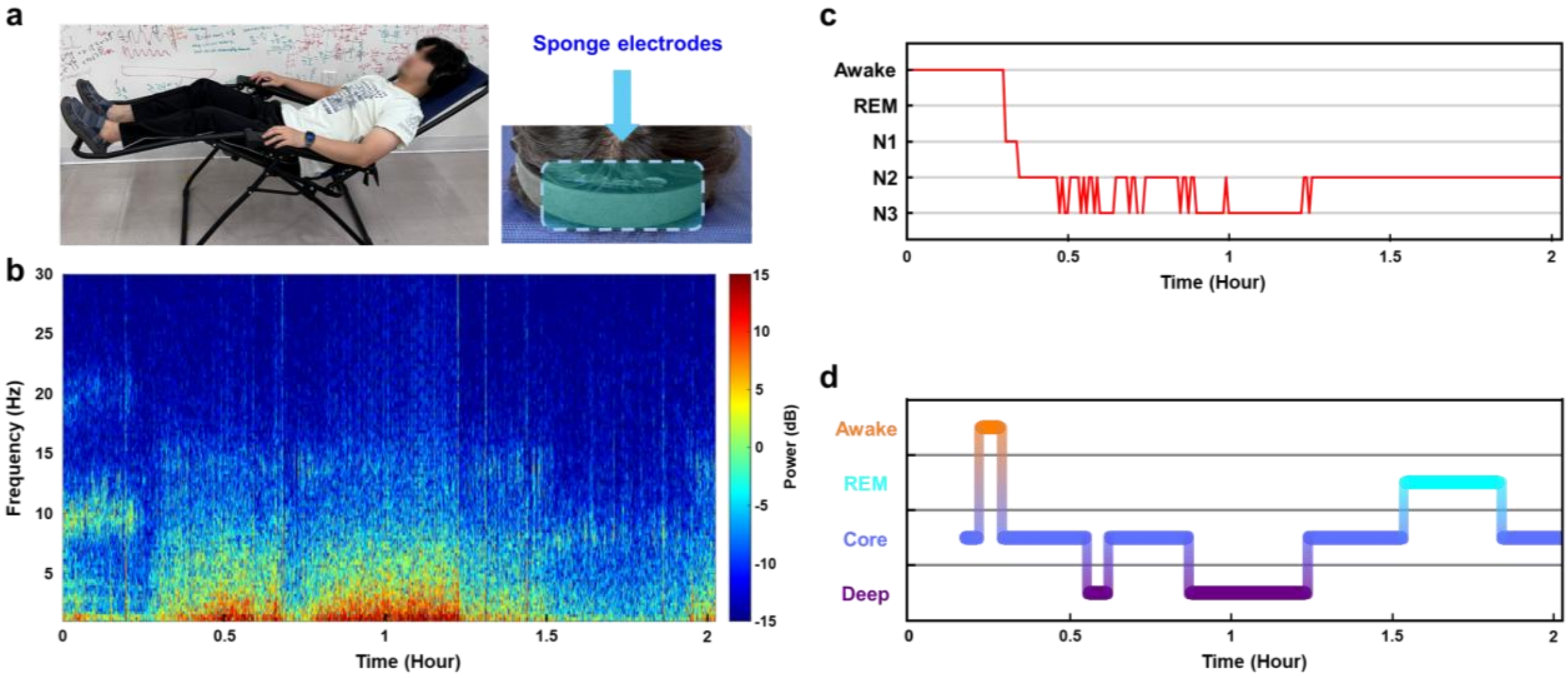

**Figure S8.** Wireless sleep EEG recording using the headphone form factor. a) Photograph of the recording setup and electrode placement under the headphones at the crown region of the scalp. b) Spectrogram of the recorded EEG. c) Hypnogram showing sleep stage transitions estimated from the EEG recording. d) Sleep stage information obtained simultaneously from the Apple Watch.

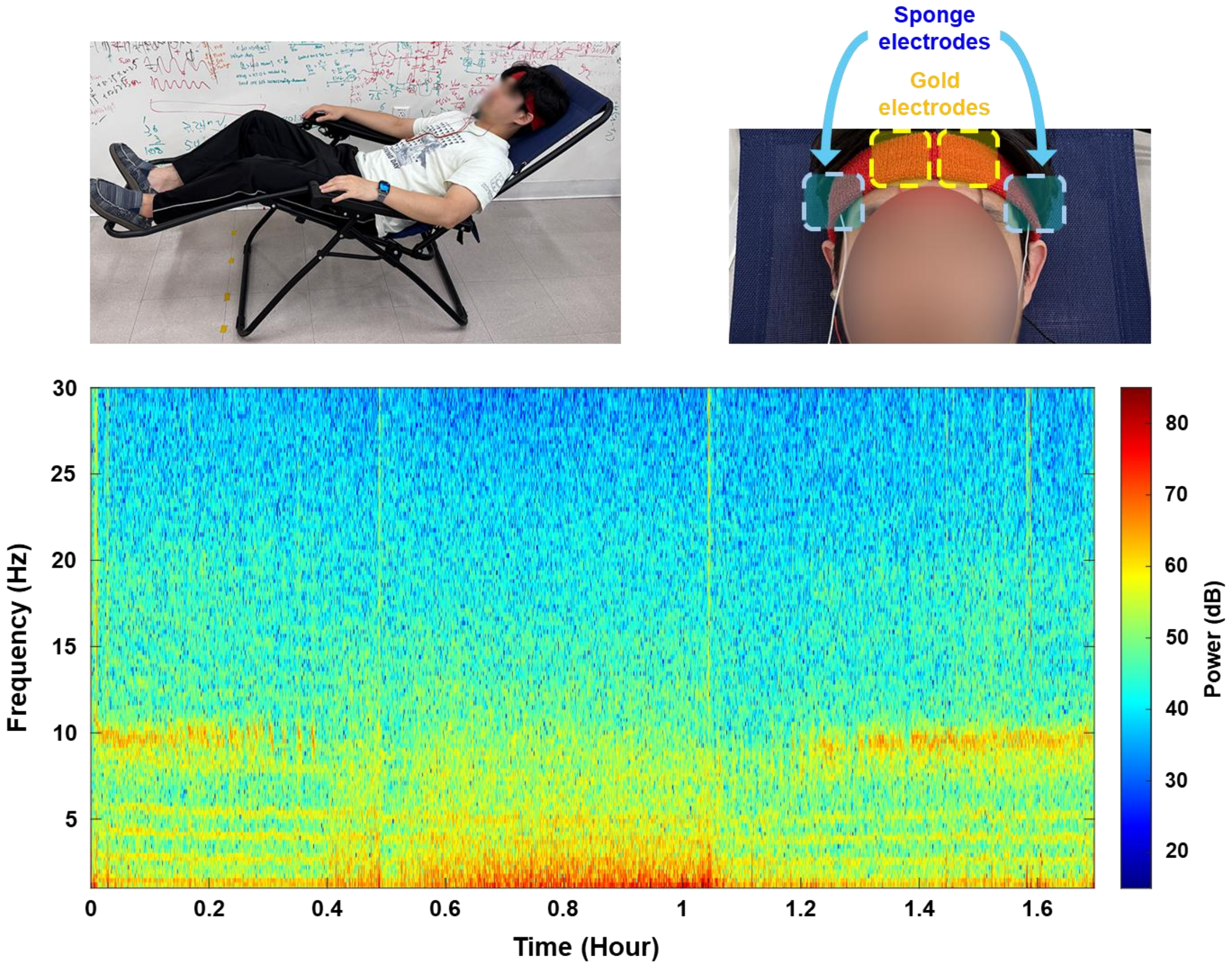

**Figure S9.** Sleep EEG recording using the OpenBCI system. Two gold cup electrodes were placed on the forehead for cross validation, while two sponge pillar electrodes were positioned on the hairy scalp near the temples.

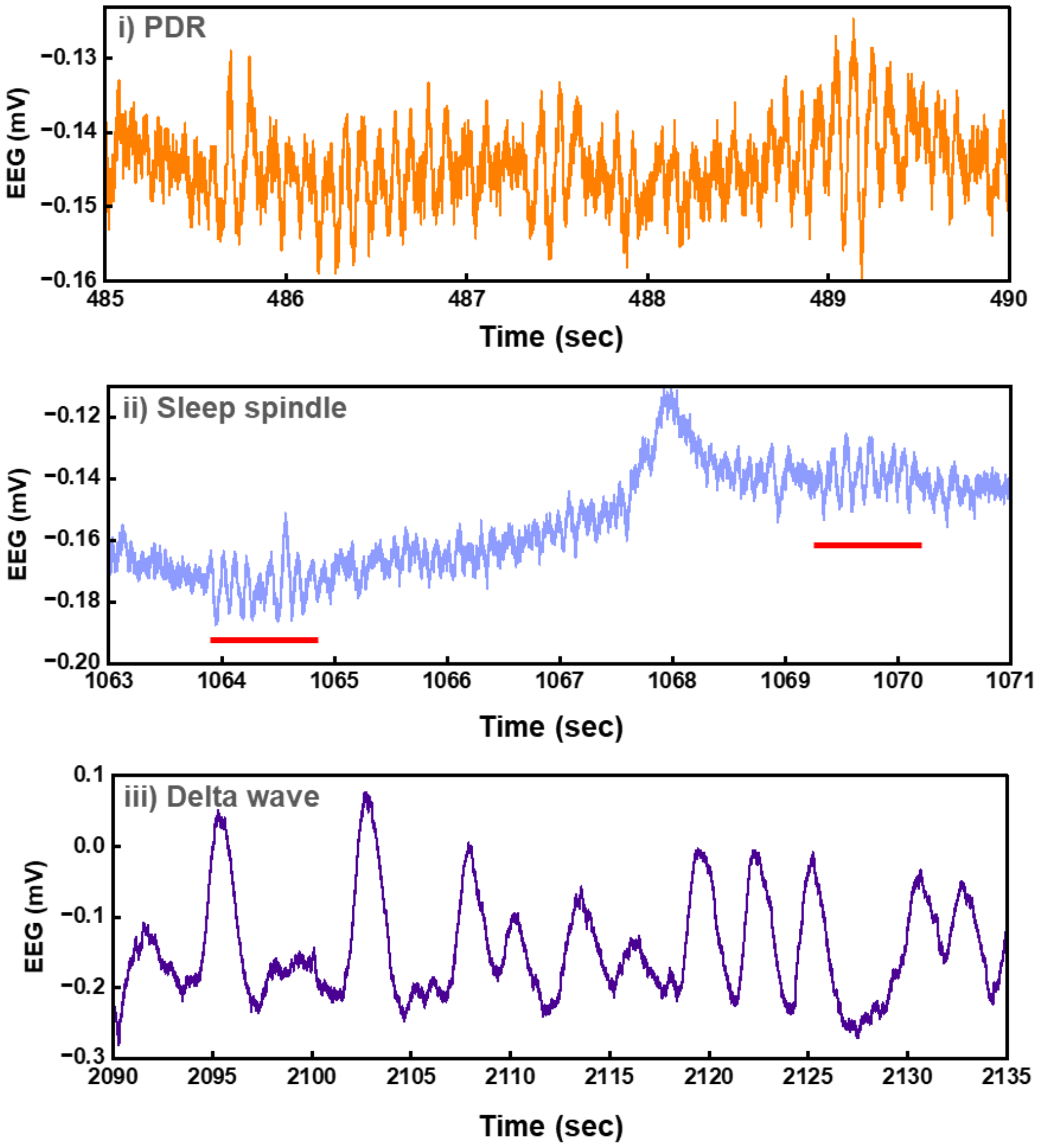

**Figure S10.** Representative sleep EEG signals. i) Posterior dominant rhythm (PDR), observed during wakefulness, showing rhythmic alpha activity. ii) Sleep spindles, underlined in red, observed during N2 sleep. iii) Delta waves, observed during N3 sleep, appearing as high-amplitude, low-frequency activity. These representative traces highlight canonical EEG features associated with awake, light NREM sleep, and deep NREM sleep states. All features were identified from unfiltered raw EEG recordings.